\documentclass{aa}  

\usepackage{ulem}

\usepackage{graphicx}
\usepackage{txfonts}
\begin{document}

   \title{The root of a comet tail -- {\it Rosetta} ion observations at comet 67P/Churyumov--Gerasimenko}

   \author{E. Behar
          \inst{1,2},
          H. Nilsson
          \inst{1,2},
          P. Henri
          \inst{5},
          L. Ber\v ci\v c
          \inst{1,2},
          G. Nicolaou
          \inst{1,4},
          G. Stenberg Wieser
          \inst{1},
          M. Wieser
          \inst{1},
          B. Tabone
          \inst{7},
          M. Saillenfest
          \inst{6},
          C. Goetz
          \inst{3}}
   \authorrunning{E. Behar et al.}
          
   \institute{Swedish Institute of Space Physics, Kiruna, Sweden.\and
              Lule\aa\ University of Technology, Department of Computer Science, Electrical and Space Engineering, Kiruna, Sweden.\and
              Technische Universit\"{a}t Braunschweig, Institute for Geophysics and Extraterrestrial Physics, Mendelssohnstra\ss e 3, D-38106 Braunschweig, Germany.\and
              Mullard Space Science Laboratory, University College London, Dorking, UK\and
              LPC2E, CNRS, Orl\'eans 45071, France.\and
              IMCCE, Observatoire de Paris, PSL Research University, CNRS, Sorbonne Universit\'e, UPMC Univ. Paris 06, LAL, Universit{\'e} de Lille, 75014 Paris, France\and
              LERMA, Observatoire de Paris, PSL Research University, CNRS, Sorbonne Universit\'e, UPMC Univ. Paris 06, 75014 Paris, France
              \email{etienne.behar@irf.se}             }

   \date{Received 2018-02-16; accepted 2018-04-26}

% \abstract{}{}{}{}{} 
% 5 {} token are mandatory
 
  \abstract
   {The  first 1000 km of the ion tail of comet
67P/Churyumov--Gerasimenko were explored by the European {\it Rosetta} spacecraft, 2.7 au away from the Sun.}
   {We characterised the dynamics of both the solar wind and the cometary ions on the night-side of the comet's atmosphere. } % and characterising it in the perspective of [usual terminator plane]
   {We analysed {in situ} ion and magnetic field measurements and compared the data to a semi-analytical model.}
   {The cometary ions are observed flowing close to radially away from the nucleus during the entire excursion. The solar wind is deflected by its interaction with the new-born cometary ions. Two concentric regions appear, an inner region dominated by the expanding cometary ions and an outer region dominated by the solar wind particles.}
   {The single night-side excursion operated by {\it Rosetta} revealed that the near radial flow of the cometary ions can be explained by the combined action of three different electric field components, resulting from the ion motion, the electron pressure gradients, and the magnetic field draping. The observed solar wind deflection is governed mostly by the motional electric field $-\mathbf{u}_{ion} \times \mathbf{B}$.}


   \maketitle
   
\section{Introduction}
        
        All particles originating from a comet's nucleus and subsequently ionised by solar radiation or electron impact eventually escape the comet, reaching on average the velocity of the solar wind. Because of the fast relative speed between the comet and the solar wind, the escaping ions are collimated into a narrow tail, known as the ion tail or gas tail. Contrary to the dust tail, this plasma structure can emit light by fluorescence, and depending on the conditions can be seen from Earth. The shape of these visible  ion tails gives a major clue to the existence of both the solar wind and the interplanetary magnetic field (IMF), as for instance elaborated by \citet{alfven1957tellus}. To that extent, the ion tail is one of the major aspects of the interaction between the solar wind and the comet's atmosphere (coma).\\
        
        During the first part of the 20th century, comets played an important role in the development of space physics. The first suggestion of the existence of a steady stream of charged particles flowing away from the Sun can be attributed to Arthur Eddington \citep{durham2006nr, eddington1910mnras}, based on observations of comet Morehouse and the analysis of the shapes of the envelopes seen in the atmosphere and in the ion tail. Another model presented by \citet{biermann1951za} describes the tail of comets being dragged by a corpuscular radiation emitted by the Sun. In this description, the momentum is transferred from the wind to the tail by Coulomb collisions. During the same decade,  another major model was proposed by \citet{alfven1957tellus} where the magnetic field of the Sun is `frozen' in a similar flow of solar particles. The local addition of  a cold and slow cometary ion population would correspond to a significant decrease of the total plasma velocity, and the {frozen-in} magnetic field piles up and drapes itself around the dense coma. This draping pattern was given as an interpretation of the typical pattern of streamers in a comet tail. Comets have acted as
 natural solar wind probes before the first {in situ} measurements and the space exploration era because of their high production rate (with values ranging from $10^{25}$ to several $10^{30}$ s$^{-1}$) and their very low gravity: they were the perfect tracers to make the solar wind "visible" from Earth.
        
        The first {in situ} investigation of the interaction between the solar wind and a coma was conducted by the International Cometary Explorer ({\it ICE}, launched as {\it ISEE-3}, see e.g. \citealt{smith1986science}) which encountered comet 21P/Giacobini--Zinner in September 1985. Since that date, ten other probes have visited eight different comets. {\it ICE}, {\it Deep Space 1}, and probably {\it Giotto} for its second encounter actually flew through the tails of their respective comets,  21P/Giacobini--Zinner, 19P/Borelli, and 26P/Grigg--Skjellerup,  validating some of the theoretical results such as the draping of the IMF remarkably observed by {\it ICE} and presented in \citet{slavin1986grl}. Another noteworthy {in situ} result is the observation of the tail of comet Hyakutake (C/1996 B2) by the {\it Ulysses} spacecraft, 3.8 astronomical units (au) away from the comet's nucleus, reported by \citet{jones2000nature}. However, these few events are tail crossings; in other words, they only gave a snapshot of the tail structures along one straight line along the spacecraft trajectory. If the coverage along this line is optimum (from outside of the tail to inside to outside again), the spatial coverage along the radial dimension for instance is almost non-existent. The situation for one of the AMPTE mission experiments, the so-called artificial comet, was quite different to these fast passages, and allowed for a more thorough study of how momentum and energy were exchanged between the background incident plasma and the injected heavy ions \citep{valenzuela1986nature, haerendel1986nature, coates2015jp}. \\
        
\begin{figure}
\includegraphics[width=1\linewidth]{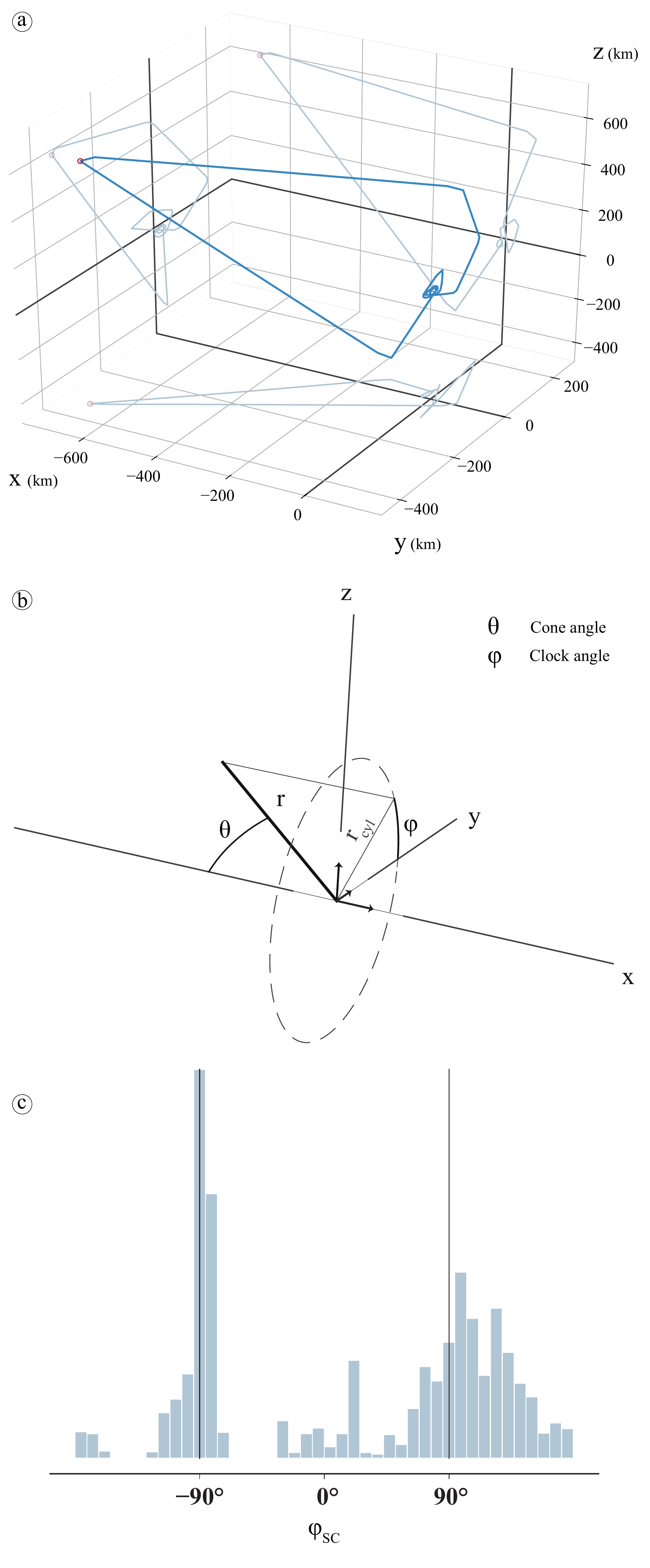}
\label{spherSys}
\caption{(a) The trajectory of the spacecraft in the CSEQ reference frame (blue) and its projections (grey). (b) The cartesian and spherical sets of coordinates. (c) The distribution of spacecraft clock angle during the period of interest, in the CSEQ frame. \label{orb}}
\end{figure}

        At the end of the 22 March 2016, as comet 67P/Churyumov--Gerasimenko (67/CG) was 2.64 au  from the Sun and orbiting away from it, the European probe {\it Rosetta} operated the first manoeuvre that would bring it on an excursion within the previously unexplored night-side of the coma. The excursion lasted for more than 14 days, during which  the spacecraft reached a maximum cometocentric distance of 1000 km, along a complex trajectory (see Figure \ref{orb}, upper panel). This was the second and last excursion operated by {\it Rosetta}; the first was  a day-side exploration conducted in October 2015 during a  different activity level and closer to the Sun. An overview of the ion data during the entire mission is shown in \citet{nilsson2017mnras}, where it can also be seen how the ion data from the tail excursion stands out in relation to all the other data.
        
        Contrasting with the otherwise low altitude terminator orbit, this tail excursion allowed the study of the root of the comet's ion tail, and specifically in our case, the study of the plasma dynamics of this region. 67P/CG is less active than most of the previously visited or observed comets (the activity is typically quantified by the production rate of neutral elements). As pointed out by \citet{snodgrass2017ptrsl}, its ion tail was actually never observed from the ground. To that extent, the interaction between the solar wind and the coma is remarkably different to what is usually described for more active comets closer to the Sun. Instead of the solar wind being deflected symmetrically around the coma and not flowing in the inner region close to the nucleus, we see a coma entirely permeated by the solar particles. By characterising the phase space distribution functions of both the cometary ions and the solar wind, we can understand how the two populations interact in this region, and to what extent this interaction takes part in the escape of cometary ions through the tail, for such a low activity comet far from the Sun. In a statistical study, \citet{bercic2018aa} recently described the cometary ion dynamics at the same heliocentric distance, close to the nucleus and in the terminator plane. Two different cometary ion populations were reported, one which  gained its energy mostly through its interaction with the solar wind upstream of the nucleus, and one that was accelerated in a region they dominate. That study is put into the perspective of this excursion, completing the larger picture of the interaction. Additionally, \citet{volwerk2018aa} present an extensive analysis of the magnetic field observed during the same excursion. The authors found that the magnetic field was not draped around the nucleus in a classical sense; instead, the magnetic field direction was mostly aligned with the IMF direction expected from a Parker spiral model. However, the tail clearly showed a two-lobed structure with regard to the wave activity: directly behind the nucleus the so-called singing comet waves \citep{richter2015ag} were very prominent, whereas to the sides their contribution to the power spectral density becomes negligible.

%__________________________________________________________________

%\section{Method}

%\subsection{Instrument description}

\section{Instrument and methods}

        {\large {\bf Instrument --}} The particle data used in this work were produced by the Ion Composition Analyzer (ICA), part of the {\it Rosetta} Plasma Consortium (RPC, \citet{nilsson2007ssr, carr2007ssr}). This instrument is an ion mass-energy spectrometer and  imager, aimed at studying the interaction between the solar wind and positive cometary ions at comet 67P/CG. The RPC-ICA data consist of count rates given in five dimensions, namely time, energy per charge, mass per charge, and incoming direction (two angles). Full energy scans are produced every 12 s and full angular scans (corresponding to the full velocity space coverage of the instrument) are produced every 192 s. The energy spans from a few electron-volts up to 40 keV in 96 steps with a resolution $\frac{\delta E}{E} = 0.07$. The instrument field of view is $360 ^\circ \times 90 ^\circ$ ($azimuth\ \times\ elevation$), with a resolution of $22.5 ^\circ \times 5.0 ^\circ$. Mass is characterised through an assembly of permanent magnets and a position detection system consisting of 32 anodes, which we will refer to as mass channels. The radial position of ions on the detector plate is a function of both the mass and the energy.\\

        {\large {\bf Distribution functions \& plasma parameters --}} Solar wind ions and cometary ions are well separated in velocity space ({i.e.} in both energy and direction) and in terms of mass. This is partly illustrated in Figure \ref{spec}, which presents a spectrogram (energy versus time) where the different species are identified on the left side of the panel. When the energy separation is poorer, the direction and the mass information allow further identification (see e.g. \citet{nilsson2015science, bercic2018aa}). For this study, the species were therefore manually selected on daily time intervals.
        
        Using these manual selections, we were able to  analyse the velocity distribution functions for each species separately. By integrating the distributions, the plasma moments of order 0 and 1 were calculated, providing the density and the bulk velocity for the different species (Figs. \ref{direction} and \ref{allCyl}). The distribution functions can be seen in Figure \ref{distrib}, where red tones are used to represent solar wind protons, and blue tones for cometary ions.\\
        
        {\large {\bf Reference frames and coordinate systems --}} The bulk velocities were initially calculated in the Comet-centred Solar EQuatorial (CSEQ) frame: the $x$-axis points towards the Sun, and the $z$-axis is oriented by the Sun's north pole direction. For geometric considerations presented later on, a particular coordinate system is used to represent most of the results, namely the spherical system $(r, \theta, \varphi)$ (see  Figure \ref{orb}, middle row): $r$ is the distance between the observation point (the spacecraft position) and the nucleus, or cometocentric distance; $\theta$ and $\varphi$ are   respectively  the cone angle and the clock angle. They can be used to describe the spacecraft's position as well as the orientation of vectors. In this system, the corresponding set of coordinates in velocity space is denoted $(r_{sw}, \theta_{sw}, \varphi_{sw})$ for protons and $(r_{com}, \theta_{com}, \varphi_{com})$ for cometary ions. In Figure \ref{distrib} and Figure \ref{allCyl}, the projection in the plane ($r$, $\theta$) of the spacecraft position is given.
        
        More relevant when it comes to the plasma environment around the comet, the comet-Sun-electric field frame (CSE) also has its $x$-axis pointing towards the Sun and the $z$-axis is along the upstream electric field. However, since defined by unmeasured upstream parameters, this frame of reference is not directly available, and we can only give an estimation of the frame orientation.
        
        Another reference frame is used in the following  when working on the solar wind proton distribution functions. For each scan of the velocity space (every 192 s), a rotation around the comet--Sun line is done in order to cancel the $v_y$-component of the proton bulk velocity before integrating all distributions together, exactly as was done by \citet{behar2017mnras} and \citet{bercic2018aa}. With the assumption that protons are deflected in a plane containing the comet--Sun line and the upstream electric field (as observed close to the nucleus in \citet{behar2016grl}), this rotation leaves the upstream electric field along the $+z$-axis and the upstream magnetic field along the $+y$-axis, in which case this particular frame corresponds to the CSE frame. In the present case, however,  this assumption might not be verified, as we discuss below, and this frame can simply be considered  a proton-aligned frame.\\

        {\large {\bf Magnetic fields --}} An estimation of the upstream magnetic field is used in the next sections. It corresponds to ACE measurements at the Earth, propagated in time using  a ballistic approach accounting for the radial and longitude differences between Earth and 67P/CG, and using the solar wind velocity measured at the Earth. Magnetic field measurements at the spacecraft location are also used, and are provided by the RPC-MAG instrument. A comprehensive description is given in \citet{glassmeier2007ssr}, and a study of the night-side excursion focusing on RPC-MAG results is given by \citet{volwerk2018aa}.

\section{Results}

\begin{figure}
         \begin{center}
            \includegraphics[width=\columnwidth]{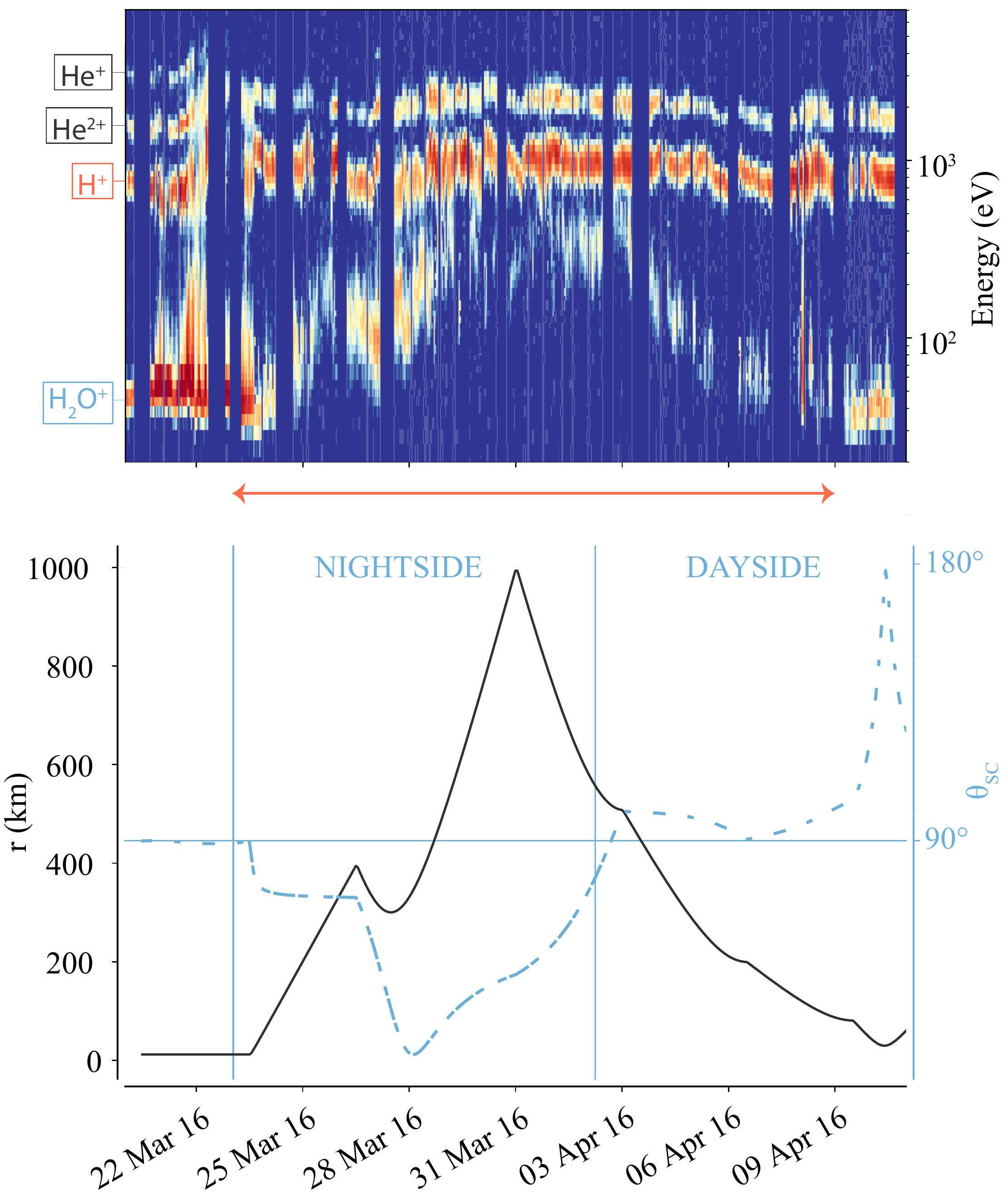}
            \caption{Upper panel:  Spectrogram of the period, with the considered period displayed by the red arrow. Lower panel:  Corresponding cometocentric distance $r$ together with the cone angle $\theta$ of the spacecraft.} \label{spec}
         \end{center}
      \end{figure}

        {\large {\bf Overview --}} The goal of this study is to characterise and compare the flow of two ion populations in the close tail environment of a comet. This requires the use of numerous combinations of coordinates in order to describe their density, their average direction and average speed, the distribution of these values,  and  their evolution through time and space as the spacecraft moves along the excursion trajectory. As seen in Figure \ref{orb}, part of the excursion is actually on the day-side of the coma, for distances below 500 km. In order to report the entire excursion, this arc is included in the analysis for distances greater than 100 km, adding up to the purely night-side region. The period over which data were analysed in this work -- from 23 March  to 8 April inclusive -- is given by the red arrow below the spectrogram in Figure \ref{spec}.
        The first combination of parameters is given in Figure \ref{spec}, upper panel, and is the combination of time, energy, and number of detections of all ions. Four different species are identified on the left-hand side of the panel, a group of three solar wind species (protons $H^+$, alpha particles $He^{2+}$, and singly charged helium particles $He^+$, which result from charge exchange between the cometary neutral atmosphere and solar wind alpha particles, see \citet{nilsson2015science}) and the cometary ions, assumed to be dominated by water molecules $H_2O^+$. The evolution of the spacecraft cometocentric distance is given in the lower panel, together with the cone angle of the spacecraft $\theta_{SC}$. One of the most striking results is already seen here, namely the correlation between the average energy of the cometary ions and the cometocentric distance. Their detection rate also seems to decrease with the cometocentric distance (as confirmed later on). In opposition, no obvious effect on the solar wind can be found in this figure (with variations in the average energy, the energy width, and the particle counts being in the usual range of fluctuations either of upstream origin or from the general interaction between the solar wind and the coma). \\

\begin{figure}
   \begin{center}
      \includegraphics[width=\columnwidth]{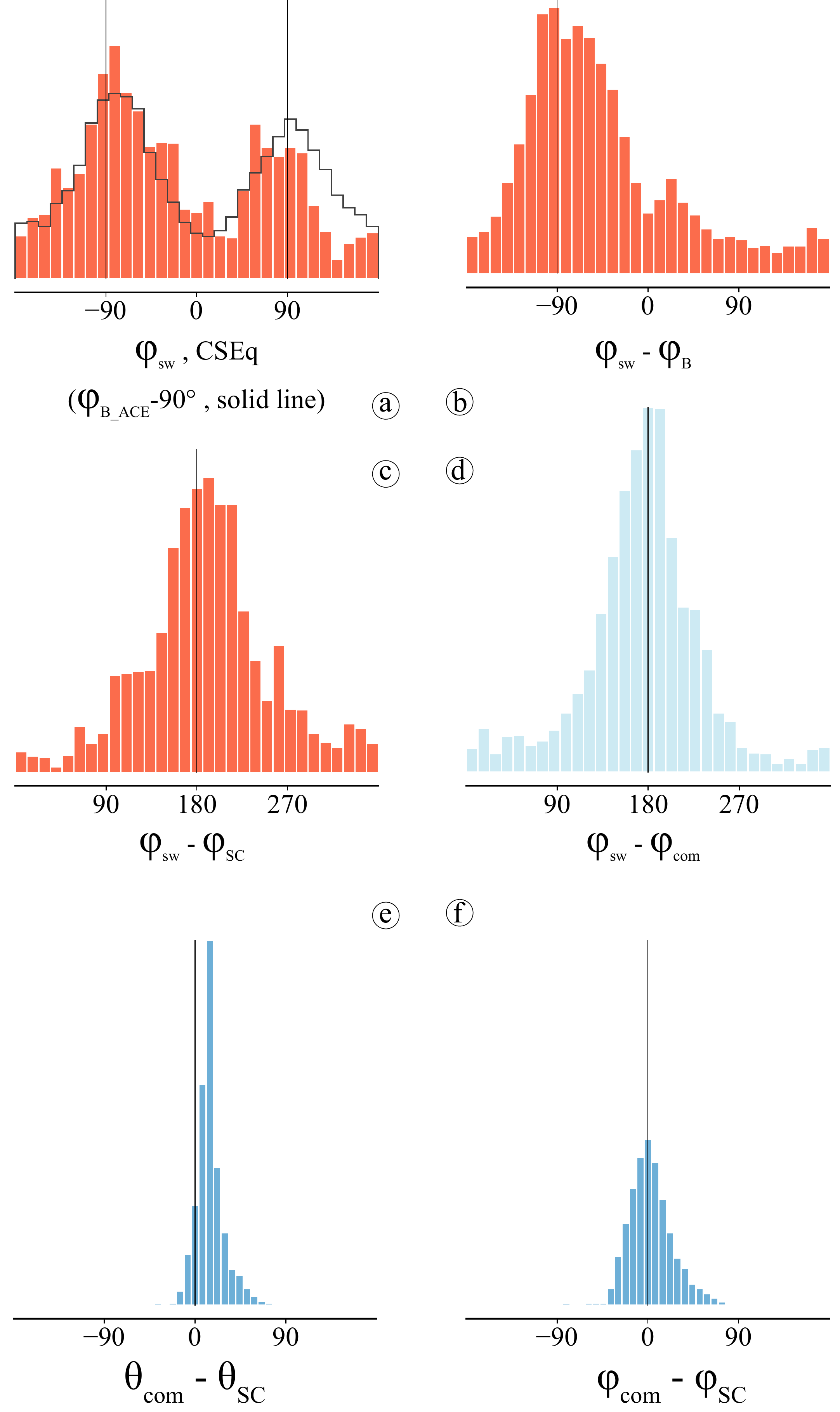}
      \caption{(a) Clock angle of the solar wind protons and the estimated upstream magnetic field in the CSEQ frame. (b) Angular difference (clock angle) between the observed magnetic field and the solar wind protons. (c) and (d) Angular difference between  the solar wind protons and the spacecraft, and the protons and the cometary ions, respectively. (e) and (f) Departure of the cometary ion flow direction from the purely radial direction for the   cone and clock angle, respectively.} \label{direction}
   \end{center}
\end{figure}

        {\large {\bf Flow directions --}} The bulk velocity directions of solar wind and cometary ions are summarised in Figure \ref{direction}. The first three histograms (a), (b), and (c) describe the proton flow direction in terms of clock angle\footnote{Only clock angles corresponding to a cone angle greater than 10$^{\circ}$ are considered to avoid poorly determined clock angles.}. In (a) they are seen deflected away from the $-x$-direction with a corresponding clock angle peaking on average at either -90$^{\circ}$ or +90$^{\circ}$ in the CSEQ reference frame. The same clock angle is on average -90$^{\circ}$ away from the magnetic field clock angle measured at the spacecraft, histogram (b), even though all configurations were observed (breadth of the distributions). The spacecraft trajectory also presents two peaks in its clock angle distribution, at -90$^{\circ}$ and +90$^{\circ}$, in Figure \ref{orb} bottom histogram, similarly to protons. In turn, the difference of the two clock angles (Figure \ref{direction} (c)) shows a very clear trend centred on 180$^{\circ}$, with 56\% of the occurrences between 135$^{\circ}$ and 225$^{\circ}$.
        
        Next, we examine the departure of the cometary ion flow from the purely radial direction, {i.e.} the direction opposite to that of the nucleus seen from the spacecraft. The difference of cone and clock angles are given by the two histograms (e) and (f), and display much narrower distributions than previously. The distribution of the cone angle difference peaks at a bit more than 10$^{\circ}$ (the particles are leaning from the radial to the anti-sunward direction), and the distribution of the clock angle difference has its maximum value at 0$^{\circ}$ on average, cometary ions were seen flowing more or less  radially away from the nucleus.
        
        The solar wind proton and cometary ion mean flow directions are compared in histogram (d). They are seen on average flowing 180$^{\circ}$ apart from each other in terms of clock angle, with once again a very broad distribution presenting all configurations. \\
        
        %In Figure \ref{interpretation}, velocity vectors for both populations are given in an estimation of the CSE frame, for positions of the space
        
        %All distributions are summing up all measurements of the entire period of interest. Using different time sub-selections (equivalent to space selection since only one trajectory was done) did not present significant changes of the distributions, and the average values as well as the width remain fairly constant. Only one change seems worth mentioning: the cometary ions become more and more radial with increasing distance to the nucleus. From a peak at around 10$^{\circ}$, the difference of cone angle shifts towards zero for distances greater than $\sim 500$ km. \\
        
\begin{figure}
   \begin{center}
      \includegraphics[width=\columnwidth]{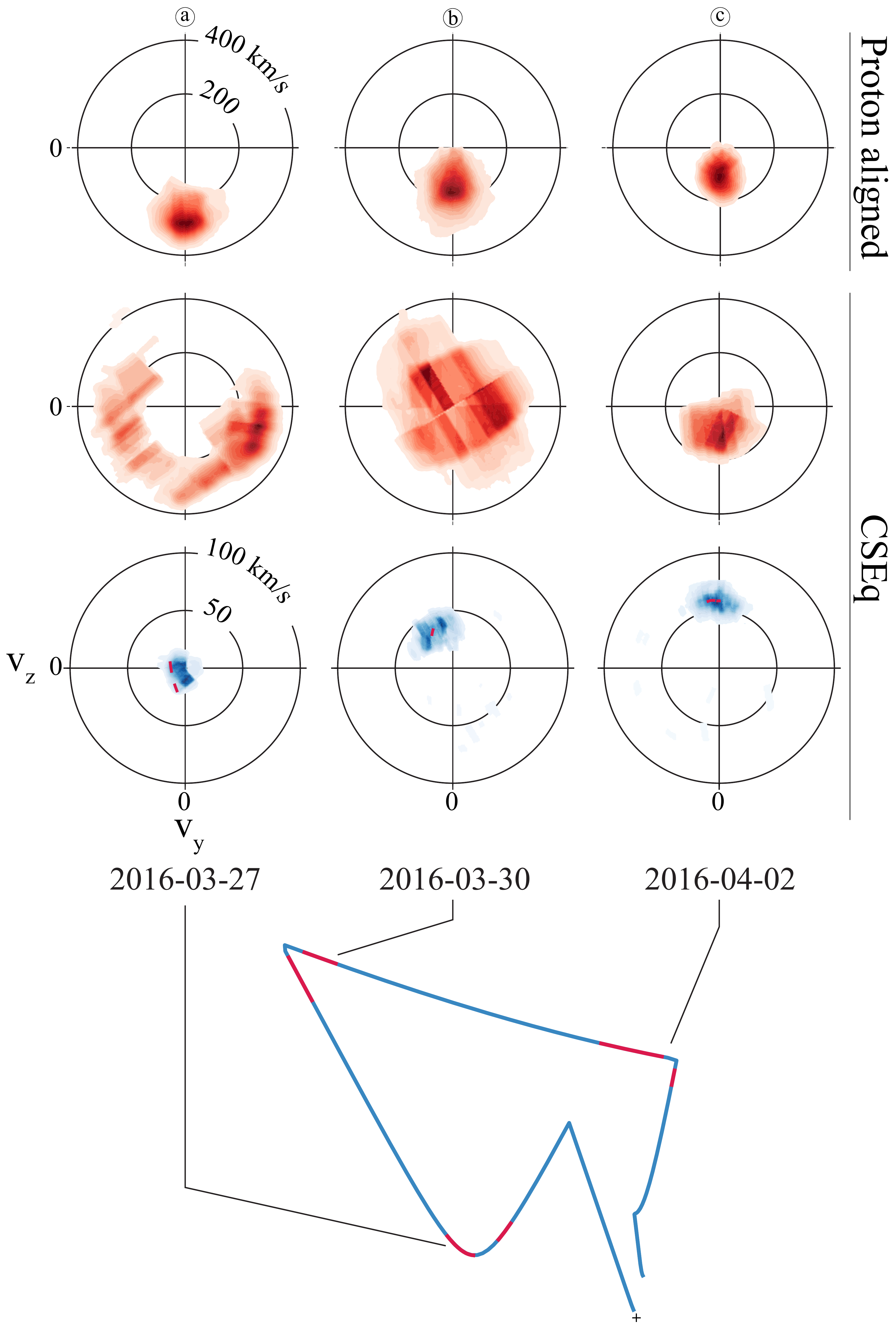}
      \caption{Distribution functions of cometary ions (blue) and solar wind protons (red) for three different dates.} \label{distrib}
   \end{center}
\end{figure}

        {\large {\bf Distribution functions --}} To go further than the bulk direction of the flows and diagnose the actual spread of the populations in velocity space, the velocity distribution functions obtained during the three different days are shown in Figure \ref{distrib}. The projection of the distributions in the plane ($v_y, v_z$) is displayed for cometary ions (blue) and solar wind protons (red) in the CSEQ frame (middle and bottom rows), and in the proton-aligned frame (top row). Data were integrated over a period of about 15 hours in each case, displayed on the orbit as the red segments (the orbit is projected in the ($r$, $\theta$)-plane). The three dates are chosen to illustrate the extreme cases: the first is taken from the inner region closest to the comet--Sun line, the second is taken from the farthest explored cometocentric distance, and the last corresponds to the region slightly day-side, close to the terminator plane, 500 km  from the nucleus. These segments are cut in two parts because of data gaps matching spacecraft manoeuvres.
        
        In the cometary ion distributions (bottom panels, blue), an indication of the radial velocity is given by the red lines: it is the velocity of particles that would flow purely radially away from the nucleus with the same average speed as the observed cometary ions. Because of the slow motion of the spacecraft, these lines are very limited.
        
        The cometary population is well focused, beam like, in the CSEQ frame, even integrated over 15 hours. The distributions are also well focused along $v_x$ (not shown here). In the same frame, the solar wind protons have much broader velocity distributions for the two first cases (a) and (b). However, the last set of distributions (Figure \ref{distrib} (c)) presents the average case seen in the histogram (d) of Figure \ref{direction}: two beams (cometary and solar wind) flowing 180$^{\circ}$ apart in the $(v_y,v_z)$-plane, with cometary ions purely radial. When aligning the bulk velocity of the solar wind protons with the $v_y$-axis for each 192 s scan, we get a well-focused beam for each day, which proves that the solar wind remains a beam at any time, only deflected and changing direction in the CSEQ frame, associated with rotations of the upstream magnetic field clock angle \citep{behar2016grl}. It was verified that the rotation also focuses the proton distributions along the $v_x$ dimension (not shown here).\\

\begin{figure*}
   \begin{center}
      \includegraphics[width=.85\textwidth]{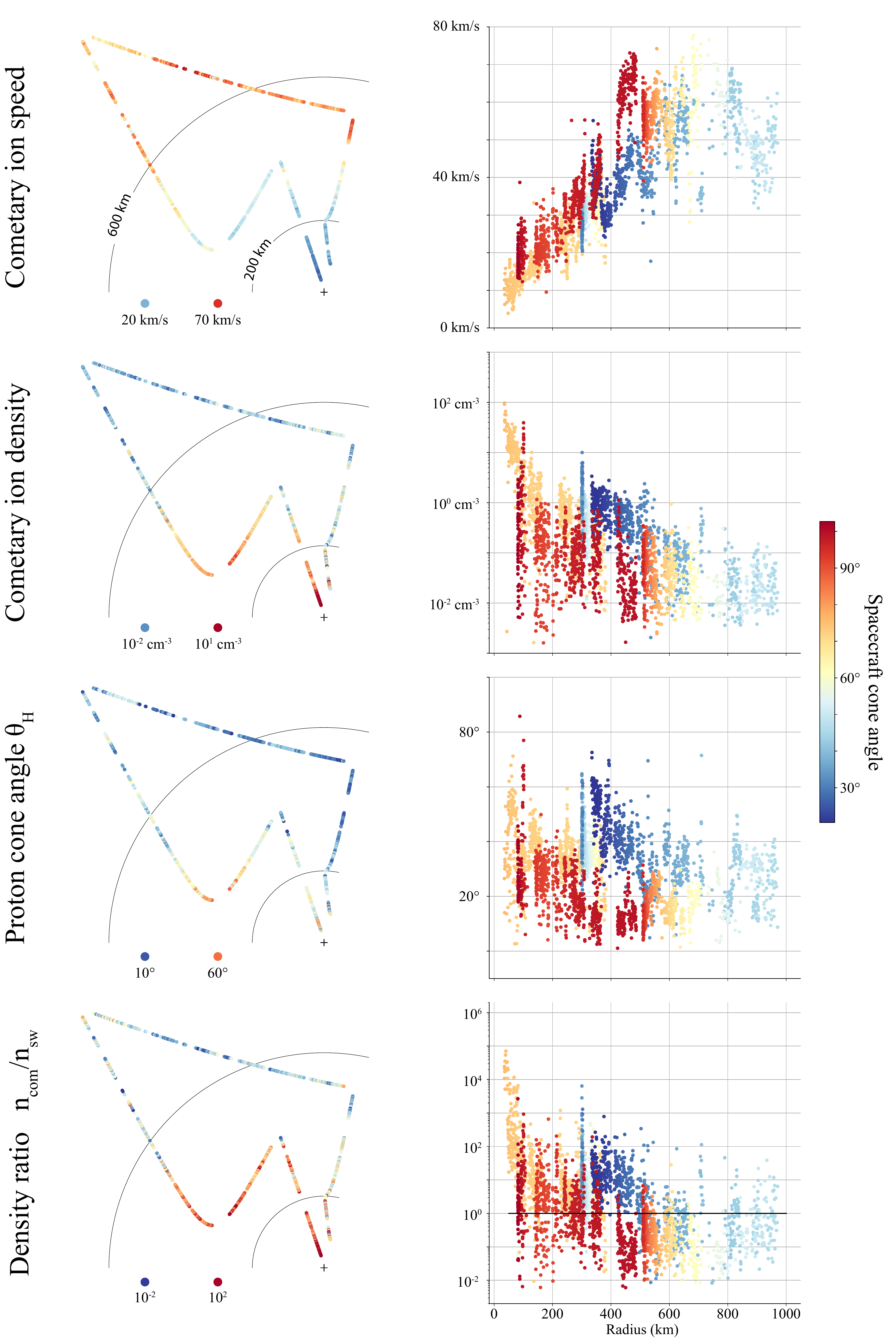}
      \caption{Plasma parameters for cometary and solar wind ions.} \label{allCyl}
   \end{center}
\end{figure*}

        {\large {\bf Cometary ion speed --}} The cometary ion speed is given in  Figure \ref{allCyl} (a) as the norm of the bulk velocity. The right-hand panel gives the evolution of the speed with the cometocentric distance $r$, and the spatial distribution measured along the spacecraft trajectory is given in the left-hand panel. The first obvious observation is that cometary ions  gain speed until a distance to the nucleus of about 600 km. Further out, the observed speed is much more spread out, and no clear trend is found. \\
        
        {\large {\bf Cometary ion density --}} The density of particles originating from the nucleus is given in Figure \ref{allCyl} (b), showing a reduction of around three orders of magnitude between closest and furthest point from the nucleus. Further away than 600 km from the nucleus, the density seems to flatten out with values between 1 and 0.1 particle/cm$^3$.\\
        
        {\large {\bf Solar wind proton deflection --}} Features of cometary origin in the solar wind speed and density could not be separated from the typical upstream variations in the observations during the excursion. However the solar wind deflection, {i.e.} the cone angle of the solar wind proton velocity $\theta_{sw}$~, shows a clear spatial evolution. In Figure \ref{allCyl} (c), we can see that for high cone angles and high radii, this deflection can be as low as 10$^\circ$, whereas for spacecraft cone angles below 45$^\circ$ and radii below 600 km the deflection reaches 50--60$^\circ$~. The distribution functions in Figure \ref{distrib} illustrate low and high deflection cases, and show how beam-like the distributions remain (in the proton-aligned frame). Such a value of 50$^\circ$ is well in line with the average deflection angles of that period, measured on the terminator plane close to the nucleus, before and after excursion. These average values can be seen in the data presented in \citet{behar2017mnras}, Figure 1, for both the alpha particles and the protons. In particular, the decrease corresponding to the night-side excursion can  be easily recognised.\\

        {\large {\bf Solar wind ions versus cometary ions --}} The charge density ratio between the cometary ions and the solar wind is given in Figure \ref{allCyl} (d) as $n_{com}/n_{sw}$~. The colour map displays a dominant cometary population as red tones and a dominant solar wind population as blue tones, whereas ratios close to 1 are displayed as pale yellow tones. Cometary ions dominate in terms of charge density 63\% of the measurements. %Not shown on this figure, we also find that for 70\% of the measurements, they have more momentum than solar wind protons (and 2 times more 60\% of the time).
        
        The density ratio plotted against the radius (right-hand panel) shows a visible correlation with the cometary ion density and the solar wind deflection. The solar wind density variations are thus much smaller than those of the cometary ion density, and do not play an important role in the evolution of this ratio. The correlation with the solar wind deflection is discussed in the next section.\\

%_________________________________________________________________

\section{Discussion}

        {\large {\bf Cometary ion origin --}} During the excursion, cometary ions are observed flowing mostly radially at all locations. It is fairly safe to believe that they originate from a region close to the nucleus. In \citet{bercic2018aa}, the authors made a statistical analysis of the cometary ion dynamics for typical orbit conditions, {i.e.} low cometocentric distance (about 30 km) and within the terminator plane. The heliocentric distance turns out to be the same for this statistical study (between 2.5 au and 2.7 au) and for the present excursion, which allows  a direct comparison between the average cometary ion behaviour at the terminator plane, close to the nucleus, and their behaviour in the night-side region. The major result of \citet{bercic2018aa} is the characterisation of two main cometary populations, namely the pick-up and the expanding populations. Whereas the pick-up ion population (average speed 30 km/s) is well ordered by the upstream solar wind electric field and originates from the day-side of the coma, the expanding population  (on average about 6 km/s) presents a cylindrical symmetry around the comet--Sun  line in terms of flow direction. However, as observed 30 km away from the nucleus, these expanding ions are not moving purely radially, and have an additional anti-sunward component. As presented by the same authors, the acceleration of this population could in part result from a radial ambipolar electric field, set up by charge separation between fast moving electrons and slower ions, due to pressure gradients of the spherically outflowing atmosphere. The cometary ions observed further out along the excursion, much faster than the expanding population, were once part of it. The extent of the region in which the ambipolar electric field may dominate has not been investigated yet, and does not appear as a sharp boundary in the present results. Far from trivial, this topic would most likely require the use of self-consistent fully kinetic numerical models. \\% They were then further accelerated in the first 600 km, with a flow direction becoming more and more radial with the cometocentric distance, in the region explored by the spacecraft.\\
        
        {\large {\bf Spacecraft position --}} On average and in the equatorial plane of the Sun, the Interplanetary Magnetic Field (IMF) is twisted in a Parker spiral, pointing either outward or inward. This average configuration corresponds to an IMF clock angle of respectively 0$^{\circ}$ or 180$^{\circ}$ in the CSEQ reference frame, which is perfectly seen in the propagated ACE data, Figure \ref{direction} (a). The corresponding upstream solar wind electric field is therefore on average along the $z$-axis, {i.e.} a clock angle of respectively -90$^{\circ}$ and +90$^{\circ}$. The deflection of the solar wind was observed during the entire mission and has been shown to be the result of the mass-loading occurring upstream of the measurement point, in a region dominated by the solar wind: the new-born cometary ions, with almost no initial velocity, are accelerated by the local motional electric field. The momentum and energy they gain is taken from the solar wind, which in turn is necessarily deflected with a clock angle opposite to that of the cometary ions (see e.g. \citet{behar2016grl, bercic2018aa}), therefore -90$^{\circ}$ away from the magnetic field clock angle. In Figure \ref{direction}, histogram (a), the propagated ACE data are shifted by -90$^{\circ}$~, and overplotted (solid black line) with the proton clock angle in the same frame. A very nice match is  found,  in shape and even in the relative hight of the two peaks. Measured at the spacecraft, the magnetic field is on average -90$^{\circ}$  from the proton clock angle as well, as seen in histogram (b).  From  histograms (a), (b), and (d) of Figure \ref{direction}, it appears that on average, the proton flow direction observed during the excursion is compatible with the effect of the mass-loading happening upstream of (and all the way to) the measurement point.
        
\begin{figure}
   \begin{center}
      \includegraphics[width=\columnwidth]{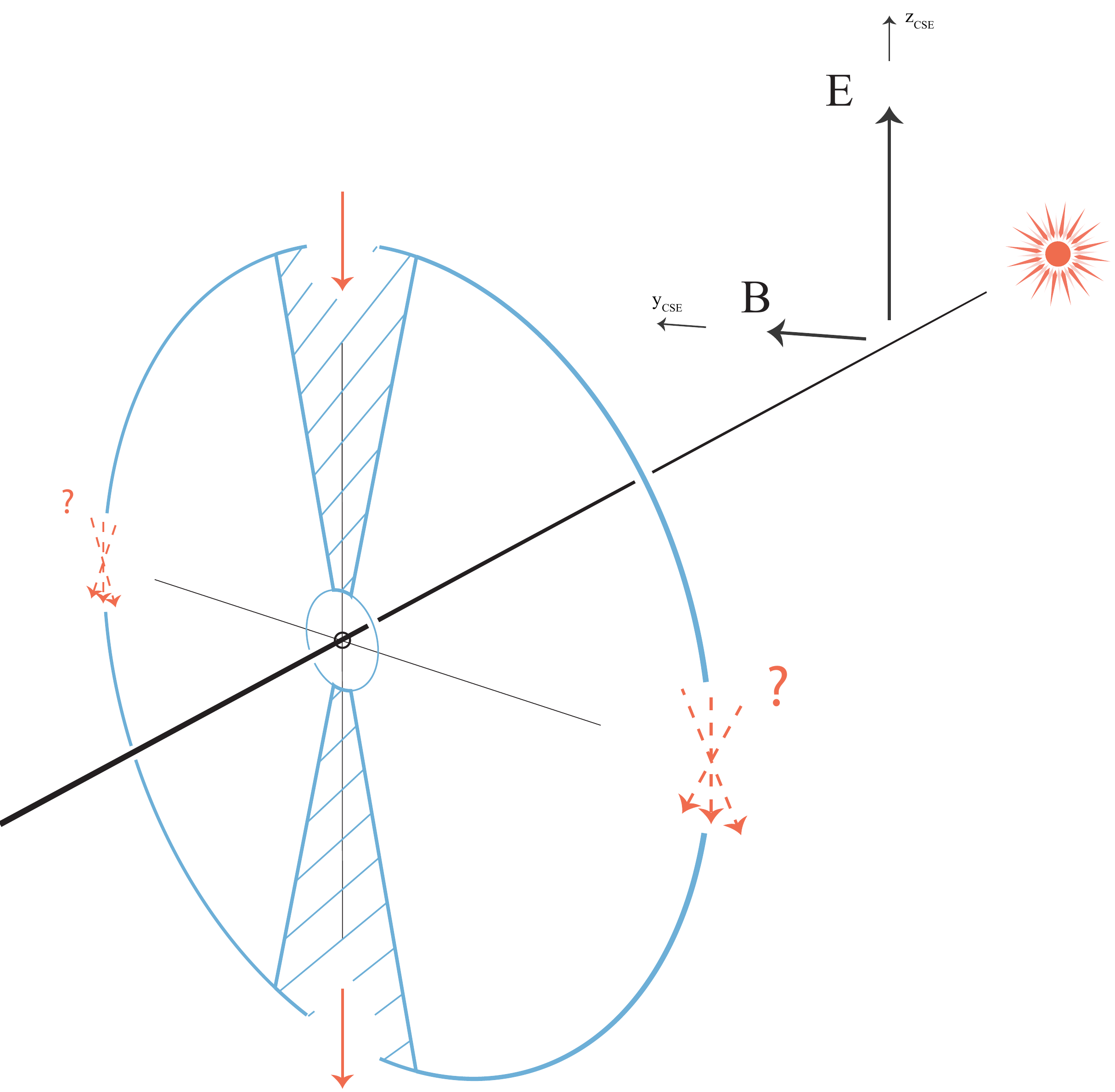}
      \caption{Representation of the CSE frame, and the regions where the selected data are estimated to be measured (striped). The red arrows indicate the expected velocity of solar wind protons, projected in the plane $(y_{CSE},z_{CSE})$. The nucleus position is not shown as it is not fixed on this representation.} \label{position}
   \end{center}
\end{figure}

\begin{figure*}
         \begin{center}
                \includegraphics[width=.8\textwidth]{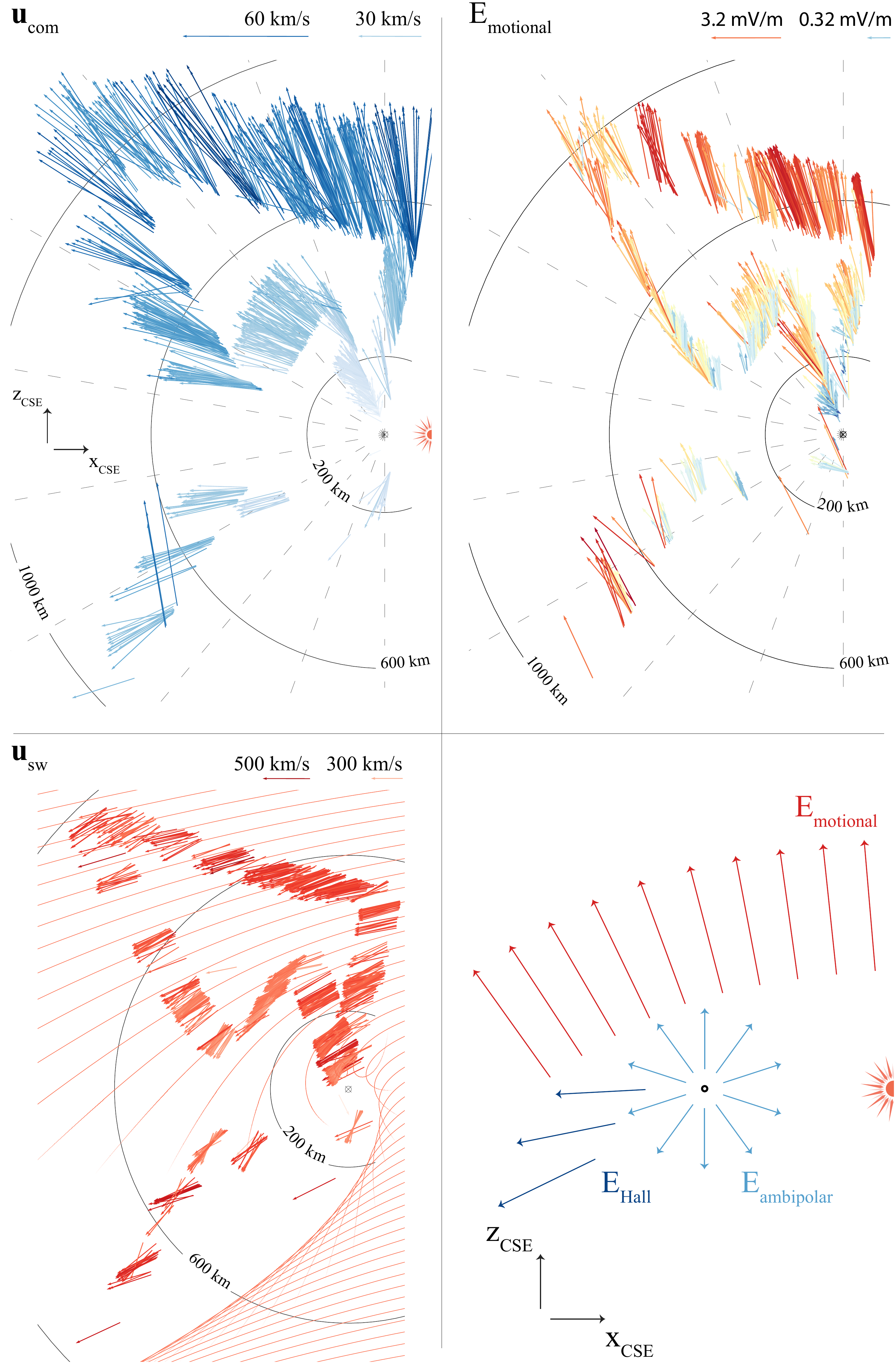}
                \caption{Upper left: Observed cometary ion bulk velocities in CSE. Lower left: Observed solar wind proton bulk velocities in CSE, with their theoretical trajectories. Upper right:  Indirectly observed motional electric field. Lower right: Schematics of the three terms of the total electric field (interpretation).} \label{interpretation}
        \end{center}
        \end{figure*}

        Assuming to first order that the solar wind protons are deflected with a clock angle of -90$^{\circ}$ {\it everywhere} in the CSE frame of reference, then a clock angle difference of 180$^{\circ}$ between the protons and the spacecraft position in CSEQ implies that the spacecraft had no $y$-component and a positive $z$-component in CSE. For a difference of 0$^{\circ}$, the spacecraft has no $y$-component and a negative $z$-component. This is illustrated in Figure \ref{position} by the two striped regions. This is the case most of the time during the excursion. However, the proton velocity can also gain a $y$-component, due to the total electron pressure gradients, magnetic field draping and upstream magnetic field cone angle (angle of the average Parker spiral). In this case, represented by dashed arrows in Figure \ref{position}, the CSE and the proton aligned frames are no longer equivalent. However, in the plane $y_{CSE}=0$ this $y$-component is close to zero because of the general symmetry of the draping pattern and the pressure (see respectively \citet{alfven1957tellus} and \citet{haser1957}). Additionally, the angle of the Parker spiral at this distance to the Sun is around 70$^{\circ}$ from the comet--Sun line, which is fairly close to 90$^{\circ}$, a value for which the ideally draped field lines are perfectly symmetric. Therefore, in summary, a clock angle difference close to 180$^{\circ}$ or 0$^{\circ}$ is still a valid indication of the spacecraft position in the CSE frame. This is illustrated in the schematic in Figure \ref{interpretation}. As seen in the previous section, almost 60\% of the observations are estimated to be within $\pm 45^\circ$ of the (y=0) plane. \\
        
        {\large {\bf The interaction --}} This subsection is extensively based on the global 2D model of the interaction between the solar wind and the coma given in \citet{behar2018aa_model}, and the analysis of its dynamics given by \citet{saillenfest2018aa}. We summarise the main aspects of the model before comparing it to the observations.\\
        
        In a cometary environment, all forces but the Lorentz force ${\bf F} = e({\bf E} + {\bf v} \times {\bf B})$ (with the electric field ${\bf E}$, the magnetic field ${\bf B}$, and the velocity ${\bf v}$ and the single charge $e$ of the considered particle), can be neglected. The alpha particles are also neglected, as well as pressure gradients, collisions, and electron inertia. Furthermore, considering spatial scales $\ell$ much greater than the ion inertial length $d_i$, the total electric field is simplified to its ideal MHD form, hereafter referred to as $\mathbf{E}_{motional}$. Considering two plasma beams (one solar wind proton beam and one cometary ion beam), we have
        \begin{equation}
        \begin{array}{rcl}
                {\bf E}_{motional} & = & -{\bf u}_{ion} \times {\bf B} \\
                                                         & = &  -\left(\dfrac{n_{sw}}{n_{sw} + n_{com}}{\bf u}_{sw} + \dfrac{n_{com}}{n_{sw} + n_{com}} {\bf u}_{com}\right)   \times {\bf B}
        \end{array}
        \end{equation}
        
        The corresponding Lorentz force on each species then reads
        
        \begin{equation}
        \begin{array}{rcl}
                {\bf F}_{sw} & = & e~\dfrac{n_{com}}{n_{sw} + n_{com}} ~ ({\bf u}_{sw}-{\bf u}_{com}) \times {\bf B} \\
                {\bf F}_{com} & = & -e~\dfrac{n_{sw}}{n_{sw} + n_{com}} ~ ({\bf u}_{sw}-{\bf u}_{com}) \times {\bf B}
        \end{array}
        \end{equation}
        
        The model then focuses on the plane $(y_{CSE}=0)$. The symmetry of this ideal system (as mentioned in the previous section) gives in this plane ${\bf B}=B_y \ \hat{{\bf y}}$~. Assuming that $u_{sw} \gg u_{com}$, we get
        
        \begin{equation}
        \begin{array}{rcl}
                {\bf F}_{sw} & = & e\ B ~\dfrac{n_{com}}{n_{sw} + n_{com}} ~ {\bf u}_{sw} \times \hat{\bf y} \\
                {\bf F}_{com} & = & -e\ B ~\dfrac{n_{sw}}{n_{sw} + n_{com}} ~ {\bf u}_{sw} \times \hat{\bf y}
        \end{array}
        \end{equation}
        
         By considering only new-born cometary ions flowing radially away from the nucleus at the speed $u_0$, and neglecting the accelerated (or pick-up) cometary ions, the dynamics of the protons is reduced to
         
         \begin{equation}
                {\bf F}_{sw} = \frac{m_{sw} \ \eta}{r^2} {\bf u}_{sw} \times \hat{\bf y} \quad \quad ; \quad \eta = \frac{e \ \nu_i Q B_{\infty}}{4 \pi \ \nu_{ml} \ n_{sw} m_{sw} u_0} \quad [m^2/s]
                \label{force}
        ,\end{equation}
        where $Q$ is the production rate of neutral elements, $\nu_i$ is the ionisation rate (taken to be constant through the coma), $B_\infty$ is the amplitude of the upstream magnetic field, and $n_{sw}$ the density of the solar wind, also assumed to be constant at zero order. The parameter $\nu_{ml}$ is a non-physical destruction rate of cometary ions, which allows us to neglect the accelerated cometary ions in the analytical expression of the cometary ion density. All the previous values are taken from the literature, and $\nu_{ml}$ is estimated to be about 0.01 s$^{-1}$ based on the values found in \citet{behar2016grl}. When compared with the data of the excursion, $\nu_{ml}$ is taken as a free parameter to allow a better fit to the data and to allow it to  absorb the uncertainty of all other parameters. However, its final value is  found to be in this precise range.
        
        As seen in Equation (\ref{force}), the solar wind protons experience a force always orthogonal to their velocity and with an amplitude proportional to $1/r^2$~. Protons are not decelerated and are only deflected. These dynamics have been thoroughly analysed by \citet{saillenfest2018aa}. The resulting proton trajectories are given in Figure \ref{interpretation}, bottom-left panel, as red lines. One characteristic of the dynamics is the formation of a caustic, along which proton trajectories intersect, a structure also observed in numerical models (see \citealt{behar2018aa_model} for a discussion of numerical models). In the resulting flow lines, two types of trajectories are to be considered with caution. After passing the caustic, the trajectories of protons are unphysical as they are not aligned with the local bulk velocity: they are expected to gyrate in a more complex manner. Additionally, the region downstream of the caustic for $z_{CSE}<0$ is poorly modelled, mostly because of the outflow from the caustic (absent in this model), and the pile-up of the magnetic field is expected to be significantly more complex than modelled in this precise region (see \citealt{behar2018aa_model} for more details). Accordingly, these two types of trajectories are lightened in Figure \ref{interpretation}.
         
        As previously discussed, many observations are estimated to be taken close to the $(y_{CSE}=0)$ plane. From  histogram (c) in Figure \ref{direction}, we have selected the data points lying in the interval $[160^\circ,200^\circ]$ (the peak of the distribution) and in the intervals $[0^\circ,20^\circ]$ and $[340^\circ, 360^\circ]$, the tips of the wings of the distribution. These data represent 33\% of all the valid data. They are estimated to have been taken within $\pm 20^\circ$ of the plane $(y_{CSE}=0)$, which we illustrate with the two striped surfaces in the schematics of Figure \ref{position}. The velocity vectors of the selected data are projected in the cartesian $(x, z)$-plane of the estimated CSE frame of reference. The proton velocity vectors are compared with the 2D model, and comparison for a value of $\nu_{ml}=0.01$ s$^{-1}$ is given in Figure \ref{interpretation}, lower left panel, in the lower range of the interval estimated in \citet{behar2018aa_model}.%for two values of $\nu_{ml}$ are given, namely $\nu_{ml}=0.01$ s$^{-1}$ and $\nu_{ml}=0.024$ s$^{-1}$~, in the lower range of the interval estimated in \citet{behar2018aa_model}. 
        
        For $z_{CSE}>0$ and $r>200$ km, a very good agreement between the observed proton bulk velocity and the modelled proton flow direction is found. The data at $z_{CSE}<0$ lie in the region poorly constrained by the model, with all data points downstream of the caustic. We note that a higher value of the rate $\nu_{ml}$ corresponding to an overall smaller deflection of the solar wind would give a better fit for data points for $r<200$ km. It appears that the model together with a value of $\nu_{ml}$ of about 0.01 s$^{-1}$ accounts for the general behaviour of the solar wind close to the plane $(y_{CSE} = 0)$ as seen during the night-side excursion.\\
        
        %For $z_{CSE}>0$ and $r>200$ km, the match between the observed proton bulk velocity and the modelled proton flow direction is more than convincing for $\nu_{ml}=0.010$ s$^{-1}$. The data at $z_{CSE}<0$ lie in the region poorly constrained by the model, with all data points but two downstream of the caustic. A higher value of the rate $\nu_{ml}$ corresponding to an overall smaller deflection of the solar wind gives a better fit for data points for $r<200$ km~. Now part of the data for $z_{CSE} < 0$ are upstream of the caustic or right on it. Despite some spread in the direction, the vectors are matching pretty well the modelled flow lines in the region. The lack of deflection for $z_{CSE}>0$ and $r>200$ km is actually expected from the model, which underestimates the deflection in this precise region by neglecting the accelerated cometary ions. If there is no reason to believe that all the parameters in equation \ref{force} have been constant for two weeks, it appears that the model together with a value of $\nu_{ml}$ of about 0.02 s$^{-1}$ account for the general behaviour of the solar wind in close to the plane $(y_{CSE} = 0$ as seen during the night-side excursion. \\
        
        The model does not solve the dynamics of the cometary ions. However, it gives us a valuable hint through the expression of the force they experience:
        \[{\bf F}_{com} = -e~\dfrac{n_{sw}}{n_{sw} + n_{com}} ~ ({\bf u}_{sw} - {\bf u}_{com})\times {\bf B}\]
        For 54\% of the selected observations, cometary ions are at least one order of magnitude slower than the solar wind protons. Therefore, ${\bf F}_{com}$ is fairly orthogonal to the proton velocity. But we can go further, and can calculate either the motional electric field or the force experienced by the cometary ions since all quantities are actually measured by RPC-ICA and RPC-MAG. The result for $\mathbf{E}_{motional}$ is given in Figure \ref{interpretation} in the upper right panel. It represents the indirect measurement of that field, independent of any assumptions. The norm of the vectors depends on the density and velocity of particles and  on the magnetic field amplitude, but the direction of the vectors only depends on the velocity vectors direction and the magnetic field direction. The RPC-ICA sensor is better at measuring the velocity of particles than their density; therefore, the direction of the vectors is better constrained than their norm. It should also  be noted, however, that  magnetic field data are not ideally constrained for most of the time interval, as explained in \citet{volwerk2018aa}. Based on Figure \ref{direction} (b), we assumed the magnetic field to be orthogonal to the protons, which makes the computed electric field amplitude an upper limit. %We now compare the observed cometary ion flow direction with the observed (and modelled) solar wind flow. With ${\bf B}$ observed to be $\Delta\varphi = 90 ^\circ$ away from the protons, ${\bf F}_{com}$ is mostly directed toward +$z_{CSE}$, outward from the nucleus, in general agreement with the cometary flow. However if not close to the plane $x_{CSE} = 0$ (vertical and orthogonal to the page), both population are not radial to each other. This departure from orthogonality does not correlate with the cometary ion speed, thus the acceleration of the cometary ions cannot be accounted by only the local motional field, and some additional electric field provide a significant acceleration of the cometary ions. \\
        
        The motional electric field results in a higher cometary ion acceleration further away from the nucleus in a region where the solar wind tends to dominate (Figure \ref{allCyl}, bottom panel). The corresponding force is aligned with the flow direction in the $z>0$ region for small $x$-coordinates. Going down in the inner tail region, the cometary ions are seen gradually departing from the field direction, to end up being completely misaligned for $z<0$~. Obviously, other sources of electric field are at work there. It is noteworthy that in this region, the singing comet waves in the magnetic field \citep{richter2015ag} were found to be stronger than further out on the night-side of the coma, as shown by \citet{volwerk2018aa}. \\
        
        {\large {\bf Hall electric field \& pressure gradients --}} The generalised Ohm's law, when neglecting collisions and neglecting the electron mass compared to the ion mass, is reduced to
        
        \begin{equation}
        \mathbf{E} = -\mathbf{u}_{ion} \times \mathbf{B} + \frac{1}{n\ e} ~\mathbf{j} \times \mathbf{B} - \frac{1}{n\ e} \mathbf{\nabla}P_e
        \end{equation}
        
        with $\mathbf{j}$ the total current, $\mathbf{\nabla}P_e$ the electron pressure gradient, and $n = n_{sw} + n_{com}$ the total ion density. We have previously investigated the first of the three terms on the right-hand side of the equation, namely the motional electric field $\mathbf{E}_{motional}$~. Neglecting the displacement current ~$\partial_t \mathbf{E}$~ in the Amp\`ere-Maxwell law, the second one -- the Hall term -- becomes $ \mathbf{E}_{Hall} = (\mathbf{\nabla} \times \mathbf{B}) \times \mathbf{B}/(\mu_0\ n\ e)$~. It can arise from the magnetic field draping around the coma, which is a source of curl for $\mathbf{B}$~. Using  Amp\`ere's law, we can find an electric field along the axis of symmetry of the draping pattern, anti-sunward. Numerical models would be appropriate for investigating further this source of electric field, as was done for example by \citet{huang2018mnras} (though at a fairly different activity level). An additional source of pile-up for the magnetic field is the asymmetric Mach cone,   the caustic of the model;  therefore, $ \mathbf{E}_{Hall} $ is expected to play a role mostly in the downstream region of the caustic, tailward.
        
        The third term, the ambipolar electric field $\mathbf{E}_{ambipolar}$~, is expected to be significant close to the nucleus. At zero order, this electric field  points radially outward from the nucleus. It might be one of the reasons why close to the nucleus the modelled proton trajectories in Figure \ref{interpretation}  depart from the observed velocity vectors. Another effect arising from the ambipolar term is a polarisation electric field due to the different motion of the cometary new-born electrons and ions. The gyroradius of the cometary ions can be much larger than the interaction region, itself larger than the cometary electron gyroradius, which induces a charge separation, and in turn a polarisation electric field. This electric field contribution is explored in the analytical work of \citet{nilsson2018aa}, and results in one additional acceleration of the cometary ions with an anti-sunward component.
        
         The three electric field are summarised in the schematic of Figure \ref{interpretation}. Above the nucleus ($z_{CSE}>0$), $\mathbf{E}_{motional}$ and $\mathbf{E}_{ambipolar}$ are essentially aligned.  Towards the inner tail region, $\mathbf{E}_{ambipolar}$  gives a tailward component to the total field compared to $\mathbf{E}_{motional}$ and eventually along the comet--Sun line, $ \mathbf{E}_{Hall}$  adds up to the tailward acceleration of the cometary ions. Overall, the cometary ion acceleration ends up being mostly radial from the combination of the three terms in the generalised Ohm's law. \\
         
         We note, however, that the apparent absence of acceleration for $r>600$ km cannot be accounted for by this interpretation. Neither the ambipolar electric field nor the Hall term is expected to provide any significant acceleration there, but as this region is dominated by a barely perturbed solar wind the motional electric field is still about a few mV/m~, as seen in the upper right panel in Figure \ref{interpretation}. An instrumental effect cannot be completely discarded, even though the energy and density ranges are nominal. The validation (or discussion) of this observation may require numerical modelling of the interaction, and we already note that the fully kinetic simulation presented by \citet{deca2017prl} give a similar result, with the energy of the cometary ions increasing and reaching a plateau in the tail region ({cf.} Figure 3 of \citealt{deca2017prl}). \\
         
         {\large {\bf Hybrid simulations --}} A qualitatively identical dynamic is seen in \citet{koenders2016aa}, Figure 14, lower and upper left panels, where the same interaction is simulated for a smaller heliocentric distance of 2.3 au~. The solar wind forms an asymmetrical and curved Mach cone similarly to the caustic given by the semi-analytical model, and the overall deflection is in great qualitative agreement with both the data and the semi-analytical model. The simulated cometary ions are seen accelerated radially away from the nucleus on the night-side of the coma as well.
         
         In another relevant simulation work, \citet{bagdonat2002jcp} describe a region of weak suppression of the magnetic field downstream of the nucleus and aligned with the comet--Sun line (\citet{bagdonat2002jcp}, Figure 4). The authors report a broad cycloidal tail leaving the nucleus orthogonal to the comet--Sun line (corresponding to our observation at $z>0$, above the nucleus), and a different cometary ion population accelerated tailward from the nucleus, corresponding to the observed cometary ion in the inner tail region. \\

\section{Summary and conclusions}

        During the night-side excursion, the cometary ion flow and the solar wind remained very directional. The cometary beam appears quasi-radial, which is interpreted as the result of its acceleration by three different electric fields of different origins: the motion of the charge carriers, the electron pressure gradients, or the magnetic field line bending. The solar wind deflection is in agreement with the analytical expression of the the magnetic field pile-up and the motional electric field, with some possible influence of the two other electric field terms close to the nucleus and in the inner tail region.
        
        A remaining open question is the apparent lack of acceleration of the cometary ions further than 700 km from the nucleus, which is not in agreement with the electric field components discussed previously. \\
        
        %Two general regions appear in the observation. In the inner region, extending to about 600 km, the cometary ions largely dominate and are accelerated to a speed of 70 km/s, while their density decreases. This gained energy is most likely furnished by the solar wind. In this inner region and close to the nucleus, the latter presents large deflection angles, reaching 60$^{\circ}$. In the outer region, no clear gain or loss of kinetic energy can be measured for the cometary ions, and their density is also evening out. The solar wind there is dominating in terms of number density, but the momentum of both ion beams is comparable.
        
        %The correlation between the number density ratio and the solar particle deflection also tells us that most of the exchange of momentum is done fairly locally, in the scale of a few hundreds of kilometres around the nucleus, as it clearly follows the evolution of the cometary ion density, a local parameter.

        The observed plasma dynamics on the night-side could enable better constraints in the comparisons between data and various models, as the area covered is much greater than usually available during the rest of the active mission. These better constraints would also allow us to direct more precisely the analysis of the simulation results as the range of parameters is significantly greater than  has been explored in the present article. As an example, an obvious follow-up of this work would be the three-dimensional mapping of the different electric field terms in the result of a numerical simulation, which would enable the study of the interaction on the flanks of the coma away from the plane $(y_{CSE} = 0)$~. \\

\begin{acknowledgements}

This work was supported  by the Swedish National Space Board (SNSB) through grants 108/12, 112/13, and 96/15.

The work at LPC2E/CNRS was supported by ESEP, CNES, and by ANR under the financial agreement ANR-15-CE31-0009-01.

We acknowledge the staff of CDDP and IC for the use of AMDA and the RPC Quicklook database (provided by a collaboration between the Centre de Donn\'{e}es de la Physique des Plasmas (CDPP) supported by CNRS, CNES, Observatoire de Paris and Universit\'{e} Paul Sabatier, Toulouse, and Imperial College, London, supported by the UK Science and Technology Facilities Council). We are indebted to the whole {\it Rosetta} mission team, Science Ground Segment, and {\it Rosetta} Mission Operation Control for their hard work making this mission possible. 
\end{acknowledgements}


\bibliography{cometLib}

\begin{thebibliography}{29}
\expandafter\ifx\csname natexlab\endcsname\relax\def\natexlab#1{#1}\fi

\bibitem[{Alfven(1957)}]{alfven1957tellus}
Alfven, H. 1957, Tellus, IX, 92

\bibitem[{Bagdonat \& Motschmann(2002)}]{bagdonat2002jcp}
Bagdonat, T. \& Motschmann, U. 2002, Journal of Computational Physics, 183

\bibitem[{Behar {et~al.}(2017)Behar, Nilsson, Alho, Goetz, \&
  Tsurutani}]{behar2017mnras}
Behar, E., Nilsson, H., Alho, M., Goetz, C., \& Tsurutani, B. 2017, Monthly
  Notices of the Royal Astronomical Society, 469, S396

\bibitem[{Behar {et~al.}(2016)Behar, Nilsson, Stenberg~Wieser, Nemeth, Broiles,
  \& Richter}]{behar2016grl}
Behar, E., Nilsson, H., Stenberg~Wieser, G., {et~al.} 2016, Geophysical
  Research Letters, 43, 1411, 2015GL067436

\bibitem[{Behar {et~al.}(2018)Behar, Tabone, Saillenfest, Henri, Deca,
  Lindkvist, Holmstr{\"o}m, \& Nilsson}]{behar2018aa_model}
Behar, E., Tabone, B., Saillenfest, M., {et~al.} 2018, Astronomy \&
  Astrophysics, (submitted)

\bibitem[{Ber{\v c}i{\v c} {et~al.}(2018)Ber{\v c}i{\v c}, Behar, Nilsson,
  Nicolau, Stenberg~Wieser, Wieser, \& Goetz}]{bercic2018aa}
Ber{\v c}i{\v c}, L., Behar, E., Nilsson, H., {et~al.} 2018, Astronomy \&
  Astrophysics

\bibitem[{Biermann(1951)}]{biermann1951za}
Biermann, L. 1951, Zeitschrift f{\"u}r Astrophysik, 29, 274

\bibitem[{Carr {et~al.}(2007)Carr, Cupido, Lee, Balogh, Beek, Burch, Dunford,
  Eriksson, Gill, Glassmeier, Goldstein, Lagoutte, Lundin, Lundin, Lybekk,
  Michau, Musmann, Nilsson, Pollock, Richter, \& Trotignon}]{carr2007ssr}
Carr, C., Cupido, E., Lee, C. G.~Y., {et~al.} 2007, Space Science Reviews, 128,
  629

\bibitem[{Coates {et~al.}(2015)Coates, Burch, Goldstein, Nilsson,
  Stenberg~Wieser, Behar, \& the RPC~team}]{coates2015jp}
Coates, A., Burch, J., Goldstein, R., {et~al.} 2015, Journal of Physics:
  Conference Series, 642, 012005

\bibitem[{Deca {et~al.}(2017)Deca, Divin, Henri, Eriksson, Markidis, Olshevsky,
  \& Hor\'anyi}]{deca2017prl}
Deca, J., Divin, A., Henri, P., {et~al.} 2017, Phys. Rev. Lett., 118, 205101

\bibitem[{Durham(2006)}]{durham2006nr}
Durham, I.~T. 2006, Notes and Records, 60, 261

\bibitem[{Eddington(1910)}]{eddington1910mnras}
Eddington, A.~S. 1910, Monthly Notices of the Royal Astronomical Society, 70,
  442

\bibitem[{Glassmeier {et~al.}(2007)Glassmeier, Boehnhardt, Koschny, K{\"u}hrt,
  \& Richter}]{glassmeier2007ssr}
Glassmeier, K.-H., Boehnhardt, H., Koschny, D., K{\"u}hrt, E., \& Richter, I.
  2007, Space Science Reviews, 128, 1

\bibitem[{Haerendel {et~al.}(1986)Haerendel, Paschmann, Baumjohann, \&
  Carlson}]{haerendel1986nature}
Haerendel, G., Paschmann, G., Baumjohann, W., \& Carlson, C.~W. 1986, Nature,
  320, 720

\bibitem[{Haser(1957)}]{haser1957}
Haser, L. 1957, Bulletin de la Classe des Sciences de l'Acad{\'e}mie Royale de
  Belgique, 43, 740

\bibitem[{Huang {et~al.}(2018)Huang, T{\'o}th, Gombosi, Jia, Combi, Hansen,
  Fougere, Shou, Tenishev, Altwegg, \& Rubin}]{huang2018mnras}
Huang, Z., T{\'o}th, G., Gombosi, T.~I., {et~al.} 2018, Monthly Notices of the
  Royal Astronomical Society, stx3350

\bibitem[{Jones {et~al.}(2000)Jones, Balogh, \& Horbury}]{jones2000nature}
Jones, G.~H., Balogh, A., \& Horbury, T.~S. 2000, Nature, 404, 574

\bibitem[{Koenders {et~al.}(2016)Koenders, Perschke, Goetz, Richter,
  Motschmann, \& Glassmeier}]{koenders2016aa}
Koenders, C., Perschke, C., Goetz, C., {et~al.} 2016, Astronomy \&
  Astrophysics, 594, A66

\bibitem[{Nilsson {et~al.}(2018)Nilsson, Gunell, Karlsson, Brenning, Henri,
  Goetz, Eriksson, Behar, Stenberg~Wieser, \& Slapak}]{nilsson2018aa}
Nilsson, H., Gunell, H., Karlsson, T., {et~al.} 2018, Astronomy \& Astrophysics

\bibitem[{Nilsson {et~al.}(2007)Nilsson, Lundin, Lundin, Barabash, Borg,
  Norberg, Fedorov, Sauvaud, Koskinen, Kallio, Riihel{\"a}, \&
  Burch}]{nilsson2007ssr}
Nilsson, H., Lundin, R., Lundin, K., {et~al.} 2007, Space Science Reviews, 128,
  671

\bibitem[{Nilsson {et~al.}(2015)Nilsson, Stenberg~Wieser, Behar, Wedlund,
  Gunell, Yamauchi, Lundin, Barabash, Wieser, Carr, Cupido, Burch, Fedorov,
  Sauvaud, Koskinen, Kallio, Lebreton, Eriksson, Edberg, Goldstein, Henri,
  Koenders, Mokashi, Nemeth, Richter, Szego, Volwerk, Vallat, \&
  Rubin}]{nilsson2015science}
Nilsson, H., Stenberg~Wieser, G., Behar, E., {et~al.} 2015, Science, 347
  [\eprint{http://www.sciencemag.org/content/347/6220/aaa0571.full.pdf}]

\bibitem[{Nilsson {et~al.}(2017)Nilsson, Wieser, Behar, Gunell, Wieser, Galand,
  Simon~Wedlund, Alho, Goetz, Yamauchi, Henri, Odelstad, \&
  Vigren}]{nilsson2017mnras}
Nilsson, H., Wieser, G.~S., Behar, E., {et~al.} 2017, Monthly Notices of the
  Royal Astronomical Society, 469, S252

\bibitem[{Richter {et~al.}(2015)Richter, Koenders, Auster, Fr\"uhauff, G\"otz,
  Heinisch, Perschke, Motschmann, Stoll, Altwegg, Burch, Carr, Cupido,
  Eriksson, Henri, Goldstein, Lebreton, Mokashi, Nemeth, Nilsson, Rubin,
  Szeg\"o, Tsurutani, Vallat, Volwerk, \& Glassmeier}]{richter2015ag}
Richter, I., Koenders, C., Auster, H.-U., {et~al.} 2015, Annales Geophysicae,
  33, 1031

\bibitem[{{Saillenfest} {et~al.}(2018){Saillenfest}, {Tabone}, \&
  {Behar}}]{saillenfest2018aa}
{Saillenfest}, M., {Tabone}, B., \& {Behar}, E. 2018, {Astronomy \&
  Astrophysics}, {submitted}

\bibitem[{Slavin {et~al.}(1986)Slavin, Smith, Tsurutani, Siscoe, Jones, \&
  Mendis}]{slavin1986grl}
Slavin, J.~A., Smith, E.~J., Tsurutani, B.~T., {et~al.} 1986, Geophysical
  Research Letters, 13, 283

\bibitem[{Smith {et~al.}(1986)Smith, Tsurutani, Slavin, Jones, Siscoe, \&
  Mendis}]{smith1986science}
Smith, E.~J., Tsurutani, B.~T., Slavin, J.~A., {et~al.} 1986, Science, 232, 382

\bibitem[{Snodgrass {et~al.}(2017)Snodgrass, A{\textquoteright}Hearn, Aceituno,
  Afanasiev, Bagnulo, Bauer, Bergond, Besse, Biver, Bodewits, Boehnhardt,
  Bonev, Borisov, Carry, Casanova, Cochran, Conn, Davidsson, Davies,
  de~Le{\'o}n, de~Mooij, de~Val-Borro, Delacruz, DiSanti, Drew, Duffard,
  Edberg, Faggi, Feaga, Fitzsimmons, Fujiwara, Gibb, Gillon, Green, Guijarro,
  Guilbert-Lepoutre, Guti{\'e}rrez, Hadamcik, Hainaut, Haque, Hedrosa, Hines,
  Hopp, Hoyo, Hutsem{\'e}kers, Hyland, Ivanova, Jehin, Jones, Keane, Kelley,
  Kiselev, Kleyna, Kluge, Knight, Kokotanekova, Koschny, Kramer,
  L{\'o}pez-Moreno, Lacerda, Lara, Lasue, Lehto, Levasseur-Regourd, Licandro,
  Lin, Lister, Lowry, Mainzer, Manfroid, Marchant, McKay, McNeill, Meech,
  Micheli, Mohammed, Mongui{\'o}, Moreno, Mu{\~n}oz, Mumma, Nikolov, Opitom,
  Ortiz, Paganini, Pajuelo, Pozuelos, Protopapa, Pursimo, Rajkumar,
  Ramanjooloo, Ramos, Ries, Riffeser, Rosenbush, Rousselot, Ryan, Santos-Sanz,
  Schleicher, Schmidt, Schulz, Sen, Somero, Sota, Stinson, Sunshine, Thompson,
  Tozzi, Tubiana, Villanueva, Wang, Wooden, Yagi, Yang, Zaprudin, \&
  Zegmott}]{snodgrass2017ptrsl}
Snodgrass, C., A{\textquoteright}Hearn, M.~F., Aceituno, F., {et~al.} 2017,
  Philosophical Transactions of the Royal Society of London A: Mathematical,
  Physical and Engineering Sciences, 375

\bibitem[{Valenzuela {et~al.}(1986)Valenzuela, Haerendel, F{\"o}ppl, Melzner,
  Neuss, Rieger, St{\"o}cker, Bauer, H{\"o}fner, \&
  Loidl}]{valenzuela1986nature}
Valenzuela, A., Haerendel, G., F{\"o}ppl, H., {et~al.} 1986, Nature, 320, 700
  EP

\bibitem[{Volwerk {et~al.}(2018)Volwerk, Goetz, Richter, Delva, Ostaszewski,
  Schwingenschuh, \& Glassmeier}]{volwerk2018aa}
Volwerk, M., Goetz, C., Richter, I., {et~al.} 2018, Astronomy \& Astrophysics,
  1

\end{thebibliography}
\bibliographystyle{aa}

\end{document}